\documentstyle[11pt,aaspp4]{article}

\lefthead{Wandel, Peterson, and Malkan}
\righthead{Central Masses of AGN}

\def\lg{{\rm log}}
\def\bh{black hole}

\def\ad{accretion~disk~}
\def\el {emission-line}
\def\bhs{black~holes}

\def\ph{photoionization~} 
\def\ed {Eddington~}
\def\ers{{\rm erg/sec}}
\def\ms{M_{\odot}}
\def\et{et al.\ }
\def\sw{Schwarzschild~}

\def\rev{reverberation~}
\def\vFWHM{\ifmmode v_{\mbox{\tiny FWHM}} \else
            $v_{\mbox{\tiny FWHM}}$\fi}
\def\kms{\ifmmode {\rm km\ s}^{-1} \else km s$^{-1}$\fi}
\def\ers{\ifmmode {\rm erg\ s}^{-1} \else erg s$^{-1}$\fi}

\begin{document}
\hfill updated \today

\vskip 0.1in

\title{Central Masses and Broad-Line Region Sizes of Active Galactic Nuclei: 
I. Comparing the Photoionization and Reverberation Techniques }

\author{A. Wandel\altaffilmark{1,2},
B.M. Peterson \altaffilmark{3}, and
M.A. Malkan \altaffilmark{1}
}
\altaffiltext{1}{Astronomy Department, University of California, Los Angeles, CA 90095-1562,
E-mail: wandel@astro.ucla.edu, malkan@astro.ucla.edu}
\altaffiltext{2}{Permanent address: Racah Institute, The Hebrew University, Jerusalem 91904, Israel}
\altaffiltext{3}{Department of Astronomy, The Ohio State University, 
174 West 18th Avenue, Columbus, OH 43210-1106,
E-mail: peterson@astronomy.ohio-state.edu}

\begin {abstract}
The masses and emission-line region sizes of AGNs can be measured by
``reverberation-mapping'' techniques and we use these results to calibrate
similar determinations made by photoionization models of the AGN line-emitting
regions.  Reverberation mapping uses the light travel-time delayed
emission-line response to continuum variations to determine the size and
kinematics of the emission-line region.  We compile a sample of 17 Seyfert 1
galaxies and 2 quasars with reliable reverberation and spectroscopy data,
twice the number available previously.  The data provide strong evidence that
the BLR size (as measured by the lag of the
emission-line luminosity after changes in the continuum)
and the emission-line width measure directly the central mass: the virial 
assumption is tested with long-term UV and optical
monitoring data on NGC 5548.  Two methods are
used to estimate the distance of the broad emission-line region (BLR) from the
ionizing source: the \ph method (which is available for many AGNs but has
large intrinsic uncertainties), and the reverberation method (which gives very
reliable distances, but is available for only a few objects). The distance
estimate is combined with the velocity dispersion, derived from the broad
H$\beta$ line width (in the \ph method) or from the variable part (RMS) of the
line profile, in the \rev -RMS method, to estimate the virial mass.  Comparing
the central masses calculated with the \rev -RMS method to those calculated
using a \ph model, we find a highly significant, nearly linear correlation.
This provides a calibration of the
\ph method on the objects with presently available reverberation data, which
should enable mass estimates for all AGNs with measured  H$\beta$ line width.  
We find that the correlation between the masses is significantly
better than the correlation between the corresponding BLR sizes calculated by
the two methods, which further supports the conclusion that both methods
measure the mass of the central \bh .  Comparing the BLR sizes given by the
two methods also enables us to
estimate the ionizing EUV luminosity $L_{ion}$ which is directly unobservable. 
Typically it is ten times the
monochromatic luminosity at 5100\,\AA ($L_v$). The Eddington ratio for the 
objects in our sample is in the range
$L_v/L_{Edd}\sim 0.001-0.03$ 
and $L_{ion}/L_{Edd}\approx 0.01-0.3$.
 
\end {abstract}


\section{Introduction}

The most fundamental characteristics of the quasar--AGN powerhouse, 
the central mass and structure, are not well understood.
The broad emission lines probably provide the
best probe of these characteristics because the broad-line region (BLR) 
is the only part of the AGN structure where we can measure
accurate Doppler motions of gas on appropriate physical scales. 
In particular, assuming the line-emitting matter is gravitationally bound, 
and hence has a near-Keplerian velocity dispersion, it is possible to 
estimate the virial central mass.

The virial assumption $v \propto r^{-1/2}$ has been directly tested using
preliminary data for only NGC 5548 (Krolik et al.\ 1991; Rokaki,
Collin-Souffrin, \& Magnan 1993). These works demonstrated the case for a
Keplerian velocity dispersion in the line-width/time-delay data, for several
broad lines in two different BLR models. The presently available data support 
these preliminary results and provide strong evidence for Keplerian velocity
dispersion in NGC 5548 (Peterson \& Wandel 1999; cf. section 4).
Although the virial assumption of Keplerian motion is essential for estimating
the mass from the BLR observed properties, the estimate may be 
approximately correct also for models in which the line-emitting gas
is not bound, such as wind or radiation-pressure induced velocities,
because the emissivity in a diverging flow decreases rapidly, and most of 
the emission would occure near the base of the flow, when the velocity is
still close to the escape velocity, which is of the order of the Keplerian
velocity (e.g. Murray \et 1998). 

The main challenges in estimating the virial mass from the emission-line 
data are 
to obtain a reliable estimate of the size of the BLR, and 
to relate correctly the line profile to the velocity dispersion in that gas.
\smallskip 
Reliable BLR size measurements are now possible through reverberation
mapping techniques (Blandford \& McKee 1982, recently reviewed by Netzer \& Peterson 1997).
The continuum/emission-line cross-correlation function measures the
responsivity-weighted radius of the BLR (Koratkar \& Gaskell 1991). 
The spectra can be combined to form
the mean spectrum and the RMS spectrum, which identifies the variable part of
the emission line. 
This method can be used only for a limited number of AGN, because it
can be applied only to variable sources and requires well-sampled 
spectra over time scales of months to years. 
Such data are difficult to obtain and exist for
fewer than two dozen AGN, mostly low- to
moderate-luminosity Seyfert galaxies.

An alternative method of estimating the virial mass is based on estimating the
BLR distance from the central mass using photoionization theory (Netzer 1990),
and the emission-line 
width as an indicator of the velocity dispersion.  This method
only assumes the line-emitting gas is gravitationally bound, but does not
depend on the specific BLR geometry or model, so it could apply to cloud
models as well as disk models (e.g., Rokaki, Boisson, \& Collin-Souffrin
1992).
 
In its simplest version, the line ratios are used to determine the
conditions in the ionized line-emitting gas, in particular the density and the
ionization parameter $U$ (the number of ionizing photons per electron)
\begin{equation}
\label {eq:u}
U=\frac {Q}{4\pi r^2 n_e c}=\frac {L_{ion}}{4\pi r^2 \bar E n_e c}, 
\end{equation}
where
$$Q = \int_{\rm 1 Ryd} \frac {L_\nu} {h\nu} d\nu $$
is the number of ionizing photons, $L_{ion}$ is the ionizing luminosity
and $\bar E=L_{ion}/Q$ is the average energy of an ionizing photon.
With an estimate of the luminosity of ionizing spectrum,
(Mathews \& Ferland 1987; Bechtold et al.\ 1987; 
Zheng \et 1997; Laor \et 1998) the eq. \ref {eq:u} can be inverted to 
give the distance of the
ionized gas from the continuum source.
Wandel (1997) has demonstrated that the BLR sizes derived even by the basic \ph
method are in good agreement with the reverberation sizes.
He also showed that using soft
X-ray data to obtain a better estimate of the ionizing flux can significantly 
improve the correlation, quantifying the error 
introduced by the uncertainty in the ionization parameter $U$ 
and the density $n_e$.

A recent compilation of the available objects with reverberation data 
(Kaspi \et 1997)
shows that the BLR size scales roughly as the square root of the luminosity, 
which agrees with the intuitive prediction of the basic \ph method, assuming 
that BLR in different AGN have similar values 
(or a narrow, luminosity-independent distribution) of the
ionization parameter and density.

In contrast to the reverberation method, 
the \ph method is indirect and the 
BLR sizes derived are subject to many intrinsic uncertainties. The 
line emission is probably extended in radius, 
and includes emission from gas with a range of
densities, so that single-zone models oversimplify the
real situation. While simple \ph models may not be trusted for
 estimating BLR parameters, it should be
possible to calibrate the \ph method using
reverberation results (Wandel 1998).
 
Since the \ph method can be used for any AGN, even
with low-resolution spectroscopy (and even with a single observation), 
it has enormous potential. If the method can be calibrated, 
as we attempt to in this work, it could give virial
masses of a large number of AGN, virtually all of those having 
reasonably well-established H$\beta$ line widths and UV line ratios.
A growing number of high-redshift quasars are also now being observed
in the H$\beta$ line with good infrared spectroscopy, which we will
also use (McIntosh \et 1998).
Such a calibration will provide us with a powerful tool to
determine reliable BLR sizes and virial masses of a large number of 
AGN, and improve our understanding of the $M/L$ relation 
and the structure of the BLR.

In the present work we investigate 
whether or not such a calibration is feasible, 
using all the presently available reverberation and emission-line
variability data.

An important feature of this paper is that we present a sample of AGN that
have been analyzed in a consistent fashion, providing reliable estimates (to
within basically geometrical factors of order unity) of AGN black hole masses.
The method of estimating virial masses is superior to what has been done
previously, as we use the scale length from reverberation techniques and 
only the variable part of the emission line to obtain a Doppler width by measuring
the line width in the rms rather than the mean spectrum. The sample analyzed
here is about twice as large as any previously discussed sample.

In the next section we describe the \ph  method. Section 3
describes the \rev - RMS method and data. In section 4, we consider the
multi-year observations of NGC 5548. Sections 5-6 give the 
correlation results for the central masses and the BLR sizes,  section
7 discusses the ionizing luminosity and \ed ratio estimates, and section 8 
discuses our result and the uncertainties associated with the sample selection, with the \rev methods and of the \ph method.

\section{The Basic Photoionization Method}

As a preliminary test we use the calibration of the
ionization-parameter method in its most basic form --- the BLR modeled by
a single thin shell, with an average value of $Un_e$ assumed for all objects. 
Combining the definition of the ionization parameter 
with the assumption of virialized velocity 
dispersion gives the simplest version of the \ph virial 
mass estimate:
\begin{equation}
\label {eq:m}
M_{ph} \approx \frac {R_{ph}v^2} {G} = K \left ( {L_{ion} \over  U n_e \bar E}
\right )^{1/2} (\vFWHM)^2 
\end{equation}
where $K=f_k G^{-1}(4\pi c)^{-1/2}$ and $f_k$ is the factor relating the 
effective velocity dispersion to the projected radial velocity deduced
from the emission-line Doppler broadening, and $R_{ph}$ is the radius
derived from eq. \ref{eq:u}.
Relating the velocity dispersion to the FWHM we assume 
$<v^2>={3\over 4}\vFWHM^2$, so $f_k=3/4$ (Netzer 1990).

Since we intend to find the normalization factor with respect to the
reverberation-rms direct method, we do not have to worry about the precise
factors, but keeping the constants can give an estimate of the effective
values of the physical parameters.  The size of the BLR is then derived from
eq.\ (\ref{eq:u}).  As we do not know the ionizing luminosity, we need to
estimate it.  In this work, as a zero-order estimate we assume it is
proportional to the visible luminosity, and try to determine the sample
average of the proportionality constant by the calibration. A priori it is
reasonable that this assumption would contribute to the errors, because
individual AGN may have different $L_{ion}/L_V$ ratios. However, as we see
below, apparently this is not the case, at least for the sample at hand.  More
elaborate estimates of the ionizing continuum may be attempted by using near
UV (Zheng \et 1997) or soft X-ray (Wandel 1997; Laor 1998) data, but it is not
clear how much these improve the estimate of $L_{ion}$, because we do not know
the shape of the EUV, and these data are also not available for many AGN; our
intention is to develop an ``affordable'' method, applicable to as many AGN as
possible.  Finally, the spectral correction enters as a square root, which
reduces the eventual errors contributed to the mass estimate.  Relating the
ionizing luminosity to some measured luminosity $L$ eq.\ (\ref{eq:u}) gives
\begin{equation}
\label {eq:r}
R_{ph}\approx 13 \left ( {f_L L_{44}\over Un_{10}\bar E_1 }\right )^{1/2}~{\rm lt-days},
\end{equation}
where $ L_{44}=L/10^{44}\,\ers $ and $f_L=L_{ion}/L$ is the factor relating
the observed luminosity $L$ to the ionizing luminosity,
$\bar E_1=\bar E /1{\rm Ryd}$
which gives
\begin{equation}
\label {eq:m7}
M_{ph} \approx (2.8~10^6\ms ) f \left ( {L_{44}\over Un_{10}}\right )^{1/2}
{v_{3}}^2,
\end{equation}
where $f=f_k f_L^{1/2}\bar {E_1}^{-1/2}$
and $v_3=\vFWHM/10^3$\,\kms.

As a first approximation, we assume the ionizing luminosity is proportional to
the visual luminosity $L_{V} \approx \nu L_\nu (5100\,{\rm \AA})$. 
For our purpose this is
equivalent to assuming that the ionizing luminosity is proportional to $\nu
L_\nu$.

As discussed in section 7 below, 
comparing the reverberation and \ph BLR sizes can
be used to estimate $f_L$.

\section {Reverberation-RMS Masses }
\subsection {The Method}
The \rev BLR sizes and the masses are derived by Peterson \et 
(1998a)\footnote{On account of several typesetting errors in the original paper,
the values for masses are superseded by the masses given in Table 2
below.}. 
The size of the BLR is measured by observing the light travel-time delayed response of the emission line to continuum variations. The relationship 
between the light curve $L(t)$ and the \el intensity $I(t)$ is assumed to be
$$I(t)=\int\Psi(\tau )L(t-\tau)d\tau ,$$
where $\Psi(\tau )$ is the transfer function (Blandford \& McKee 1982).
The transfer function depends on the BLR geometry, the viewing angle and the
line emissivity.
The cross-correlation of the continuum and emission-line 
light curves is given by
$$CC(\tau )=\int\Psi(\tau ') AC(\tau '-\tau)d\tau ' ,$$
where AC is the continuum autocorrelation function (Penston 1991).
It can be shown (Koratkar \& Gaskell 1991) that the centroid of the cross-correlation function , $\tau_{cent}$, 
gives the size $c\tau_{cent}$ associated with the emissivity-weighted mean 
radius for the emission region of a particular \el .

The root-mean-square (RMS) spectrum defined as
\begin{equation}
\sigma (\lambda )= \left [(N-1)^{-1}\sum_i^N (F_i(\lambda )-\bar F(\lambda))^2 
\right ] ^{1/2}
\end{equation}
where $N$ is the number of spectra and $\bar F$ is the average spectrum.
The RMS spectrum measures the variations about the mean, automatically 
excluding constant features such as narrow emission lines, 
Galactic absorption and
constant continuum and broad line features.
The RMS H$\beta$ emission-line profile 
gives the velocity dispersion in the variable
part of the gas, the one which is used to calculate the BLR size.

The line width and the BLR size may be combined to yield the virial
"reverberation" mass estimate (equivalent to eq. \ref{eq:m7})
\begin{equation}
\label {eq:mrev}
M_{rev} \approx (1.45\times 10^5\ms )  \left ( {c\tau\over {\rm light-days}}\right ) {v_{rms,3}}^2,
\end{equation}
where  $v_{rms,3}=\vFWHM (rms)/10^3$\,\kms.

\subsection {The Data}

Our sample consists of 19 objects, 17 Seyfert 1 galaxies and 2 quasars. This 
is the most complete compilation to date of AGNs for which reverberation data 
are available. The sample will be described in a separate paper,
and in this contribution we consider only ground-based optical data. These
data are drawn from the following sources:
\begin{enumerate}
\item {\bf The Ohio State AGN Monitoring Program.} A program of approximately
weekly observations of the H$\beta$ spectral region in bright Seyfert galaxies
was carried out with the Ohio State CCD spectrograph on the
1.8-m Perkins Telescope at the Lowell Observatory from 1988 to 1997.
Data on Akn 120, 3C 120, Mrk 79, Mrk 110, Mrk 335,  Mrk 590,
and Mrk 817 used here have been published by Peterson et al.\ (1998a).
The data used here on NGC 4051 are preliminary results based on
OSU data only, and these are currently being combined with data from
other sources.
\item{\bf The International AGN Watch.} Over the last decade, this 
consortium has carried out multiwavelength monitoring programs on a number 
of AGNs, and the data are publicly available\footnote{International 
AGN Watch can be obtained on the World-Wide Web at URL
{\sf http://www.astronomy.ohio-state.edu/$\sim$agnwatch/}.}
(Alloin et al.\ 1994; Peterson 1999). In this study, we employ the AGN
Watch optical data on 
3C 390.3 (Dietrich et al.\ 1998), 
Fairall 9 (Santos-Lle\'{o} et al.\ 1997),
Mrk 509\footnote{Supplemented by additional data from Peterson
et al.\ (1998a).} (Carone et al.\ 1996),
NGC 3783 (Stirpe et al.\ 1994),
NGC 5548 (Peterson et al.\ 1991, 1992, 1994, 1999;
Korista et al.\ 1995), 
NGC 4151 (Kaspi et al.\ 1996a), and 
NGC 7469 (Collier et al.\ 1998).

\item {\bf The CTIO AGN Monitoring Program.} 
In connection with the AGN Watch program on NGC 3783 (Stirpe et al.\ 1994)
in 1992, observations of NGC 3227 (Winge et al.\ 1995) and
IC 4329A (Winge et al.\ 1996) were also made. 
For the sake of greater internal consistency, we 
reanalyzed these spectra in the same fashion as the
Ohio State monitoring data.
\item {\bf The Wise Observatory QSO Monitoring Program.} 
A spectroscopic monitoring program on low-redshift quasars is
being carried out at the Wise Observatory. Thus far, preliminary
results on PG 0804+762 and PG 0953+414 
have been published (Kaspi et al.\ 1996b),
and the authors were kind enough to make the original
spectra available to us for this analysis.
\end{enumerate}

For each galaxy, we computed the optical continuum--H$\beta$
cross-correlation using the interpolation method of
Gaskell \& Sparke (1986) as implemented by White \& Peterson (1994),
and determined the centroid $\tau_{\rm cent}$ of the cross-correlation
function using all points within 80\% of the peak value $r_{\rm max}$.
Uncertainties in $\tau_{\rm cent}$ were evaluated using the
model-independent Monte-Carlo FR/RSS technique of
Peterson et al.\ (1998b). Spectra were combined to form average
and root-mean-square (rms) spectra, and the full-width at half maximum
of the H$\beta$ feature in the average and rms spectra was measured, 
transformed to the rest frame, and is quoted as a Doppler width
in kilometers per second.

A summary of the basic data used in this investigation is given in
Table 1. The name of the galaxy appears in column (1). 
Column (2) gives the monochromatic luminosity 
$\lambda L_{\lambda}(5100\,{\rm \AA})$ 
(ergs s$^{-1}$) of the galaxy
at a rest wavelength of 5100\,\AA. This is the mean luminosity
observed during the monitoring campaign from which the
data were drawn. In each case, this has been
corrected for Galactic extinction using the $A_{\rm B}$ values
in the NED database\footnote{ The  NASA/IPAC Extragalactic Database (NED)   
is operated by the Jet Propulsion Laboratory, California Institute   
of Technology, under contract with the National Aeronautics and Space      
Administration.}. Columns (3) and (4) give the full-width at half-maximum
of the H$\beta$ emission line in the mean and rms spectra, respectively.
Column (5) gives the time lag $\tau_{\rm cent}$ for the
response of the H$\beta$ emission line, and the associated
uncertainties, which are not symmetric about the measured value.

We expect the line width in the rms spectra to be larger than the 
line width in the average spectrum
because in the former the narrow-line component is eliminated. However
there are objects for which the FWHM of lines is is {\it narrower} 
in the rms spectrum than in the average spectrum.
One possible interpretation is that the emission-line gas with the highest
velocity dispersion and nearest to the central source, which contributes to
the far wings does not vary much, perhaps because it arises in a medium
that is optically thin to ionizing radiation (e.g., Shields, Ferland,
\& Peterson 1995).
\placetable {tbl-1}

Note that for NGC 7469 we give the lag with respect to the UV
continuum rather than relative to the continuum at 5100\,\AA.
The reason for this is that in NGC 7469, the optical continuum
variations follow those in the ultraviolet continuum by about
one day. This is the only AGN in which this effect has been
measured (Wanders et al.\ 1997; Collier et al.\ 1998;
Peterson et al.\ 1998b).

\section{NGC 5548: Evidence for the Keplerian Assumption}

The basic assumption of this model is that the line-emitting gas is
gravitationally bound, and its velocity dispersion is approximately
Keplerian.  If this is the case, we would expect that during variations of
the ionizing luminosity, the ``virial mass'', given by $M\sim G^{-1} v^2 r$
remains constant.  While the radius given by the luminosity scaling of the
\ph method (eq. \ref{eq:r}) is uncertain, the reverberation radius and the
associated velocity derived from the variable part of the line (FWHM-RMS)
provide a direct measure of the virial mass.  

The virial assumption can be tested for individual objects.
It is known that the BLR is stratified, with different emission lines
different distances from the central source. Typically different lines
also have very different line widths.
If the motions in the BLR are dominated by the gravity of the central mass, 
the Keplerian relation (\ref{eq:mrev}) applied to different lines in the
same objects must yield 
the same value for the central mass. Similarly, if the continuum and \el s
vary over long time scales (the short time scale variability is applied for
deriving the \rev BLR size), the virial mass should remain constant.
In other words,
above test states that the lags $(R)$ of different lines and at different
epochs should be anticorrelated
with the rms line width as
$R\propto \vFWHM ^{-2}.$ 

We have tested the virial assumption with the multiyear multiwavelength
data on NGC 5548 (Peterson \& Wandel 1999). The emission-line lags $\tau$
range from $\sim2$ to $\sim30$ days, and are anticorrelated with
the line widths $FHWM$, consistent with a constant value of the product 
$\tau \times (FWHM)^2$, as expected if the virial assumption is
correct. There is significant scatter around the best virial fit
to the data, however; we suspect that this is due in part to the
fact that both the lag and the line width are dynamic quantities that
are changing in response to changes in the continuum flux.
Moreover the H$\beta$ results for two years (1993 and 1995) had to
be excluded from this analysis because the rms profiles were strongly
double peaked and therefore did not yield an unambiguous line width.
Again this suggests to us that while the virial interpretation is
correct to first order, it is not the whole story.

Figure 1 shows the virial mass calculated from the reverberation radius and the line width in the rms spectrum
as a function of the FWHM-rms for each line in each year.
This way of presenting the data enables us to demonstrate that the virial
mass of all lines is consistent with a single value.
 
Fitting an $M$=const. to the
points in the $M$ vs. FWHM plane is equivalent to fitting a line 
with a slope of -2 (lag$\propto $(FWHM)$^{-2}$)
 in the lag vs. FWHM (log-log) plane 
(if the two observables were uncorrelated, the best fit would be $M\propto (FWHM)^2$).

\placefigure{fig1}

We find an average virial mass (weighted by the uncertainties in the mass)
of $(6.1\pm 2)\times 10^7\ms$.
The weighted mean has been calculated as
$<x>=\Sigma (x_i/\sigma_i^2) / \Sigma (1/\sigma_i^2)$
where $\sigma_i$ is the uncertainty in $x_i$ in the appropriate direction 
(up or down) with respect to the mean.
It is easy to see that all the points in the figure are consistent with the average value within $\sim 1\sigma$.
While the annual averages of line width and lag in the NGC 5548 data set
vary by factors of 2-3 and more than an order of magnitude, respectively,
they are anticorrelated in a manner close to the Keplerian relation above, 
so that the virial mass
remains approximately constant (within the uncertainties).

\placetable {tbl-3} 
We have calculated the goodness of the $M$=const assumption in three schemes:
\begin{itemize}
\item {An $M=\bar M$ fit in the $M$ vs. FWHM plane}
\item {A $y=-2x+a$ fit in the log(lag) vs. log(FWHM) plane
(that is, x-y plane)}
\item {The best fit $y=bx+a$ in the log(lag) vs. log(FWHM) plane}
\end{itemize}
where $\bar M$ is the uncertainty-weighted average \rev mass,
$x=$log(FWHM) and y=log($\tau$).
We have performed these fits for five data sets:
\begin{itemize}
\item {All data points}
\item {All lines and H$\beta$ represented by the weighted average of the
six years (log(FWHM)=3.71, log(lag)=1.25)}
\item {All lines from the years 1989 and 1993 (when available), i.e. all lines and H$\beta$ from 1989}
\item {Only H$\beta$ for the six years, lags from the centroid of the CC function}
\item {Only H$\beta$ for the six years, lags from the peak of CC (see below)}
\end{itemize}
The last two sets demonstrate the constancy of the virial mass for time variability 
of a single line, but evidently the small number of points
does not give very significant results. The last set may test the sensitivity
of the \rev scheme to the geometry (see below).

For all sets we have calculated the reduced
$\chi^2$ in the three schemes (table \ref{tbl-3}) 
The best fit and the $\chi^2$ in the (x,y) plane 
are calculated with weighting by lag 
uncertainties (Bevington pp. 118 and 188).
The errors in the $x-y$ plane 
were estimated from the (logarithmic, directional) standard deviation $\sigma_i$ 
for each point i,

$$\sigma_i= \log (1+\sigma_i^+/\tau_i)  \qquad{\rm if}\qquad  y_i< b x_i + a_i$$

$$\sigma_i= -\log (1+\sigma_i^-/\tau_i)   \qquad{\rm if} \qquad  y_i> b x_i + a_i$$

where ${\sigma_i}^+$ and ${\sigma_i}^-$ are the + and - uncertainties in the lag $\tau$.

Note that the $M$=const fits have lower values of $\chi^2$ 
 than in the FWHM-lag fits (in the $x-y$ plane). Although the two
are equivalent,  the latter uses only the uncertainties
in the lag, while the former uses the uncertainties in $M$, which combine
the lag and FWHM uncertainties, and is therefore more appropriate.

The best fits for the three sets with all lines are
consistent with the virial assumption (slope of -2 in the x-y plane).
On the other hand,
the H$\beta$ fits are not consistent - this is due to their 
small dynamical range, which is even reduced by the weighting, as the point
with the largest FWHM also has a much larger uncertainty in the lag than the
other points. 
Also the dynamical nature of the lag and line width, which depend on the 
constantly changing continuum flux is probably contributing to the scatter.
Given these factors and the small number of points, this result is
probably not very significant.
Note however that the $\chi^2$ tests give a reasonable probability to the
$M$=contant assumption even for the H$\beta$ sets, and certainly cannot exclude it.

For the H$\beta$ annual data we have tried also a different \rev
scheme, using the lags of the peak (rather than the centroid ) of the line-continuum CC function.
This may give an estimate of how sensitive these results are with
respect to the uncertainties of the \rev technique and its dependence
on the BLR geometry (see section 8.2).

The (uncertainty-weighted) mean of the annual virial masses derived from the H$\beta$ 
line annual averages is 
$$<M_{rev-c}>=(6.6\pm 1.6)\times 10^7 \ms \qquad {\rm H\beta ~annual, ~centroid ~lag}$$
and
$$<M_{rev-p}>=(6.6\pm 0.5)\times 10^7 \ms \qquad {\rm H\beta ~annual, ~peak ~lag}.$$

In addition to being mutually consistent,
these results are in good agreement with the value 
derived by using all the available \el s and years (altogether 14 data points), 

$$<M_{rev}>=(6.1\pm 0.4)\times 10^7 \ms \qquad {\rm different ~lines, ~centroid ~lag}.$$

\section {The Mass Correlation}

The basic procedure is quite straight forward: use rms spectra of the 
objects with available reverberation analyses, and compare the virial
mass calculated this way to the value calculated with the ionization
parameter method, using  canonical AGN values for $Un_e$ and $Q/L$ rather than
specific values for each object. 

Table 2 compares the BLR sizes and the virial masses derived by the two
methods.

\placetable {tbl-2}

We now try to find the correlation and the functional relation between the
masses determined by the two methods for the objects with reverberation data.
We expect the relation to be linear, $M_{rev}\propto M_{ph}$, but we are also
testing for a different power-law dependence.  Since the parameters that we
have approximated appear in the square root, the scatter introduced through
these approximation may be small. Since we find that the two mass estimates
are well correlated, the correlation may be used to derive a functional
dependence --- to a first approximation we assume a linear correlation, which
gives the normalization coefficient in eq.\ (\ref{eq:m7}).
Figure 2 shows the virial masses calculated by the  
\ph (eq.\ \ref{eq:m7}) and the reverberation-rms methods.

\placefigure{fig2}

Considering the unweighted data points gives $\lg (M_{rev}) = 0.82~ \lg 
(M_{ph})+1.53$ with a correlation coefficient of 0.83.

\subsection{Weighting by Measurement Uncertainties}

Since the uncertainties in $M_{rev}$ are very different for different objects,
it is important to weigh the points according to their uncertainties.
Taking into account the directed 
\footnote {Directed in the sense of using the appropriate error bar (upper
or lower) of each point with respect to the fit, and repeating the fitting
until it converged.}
uncertainties in the $M_{rev}$ 
the correlation coefficient becomes 0.89, and the best fit 
$$\lg (M_{rev}/\ms ) = b ~\lg (M_{ph}/\ms ) + a ,$$
with
$b=0.93\pm 0.07$ and  $a=0.70\pm 0.53$.

For NGC 5548 masses we have used the weighted averages of the six years.
In the weighted fit we have increased the formal uncertainty of the Mrk 590 and
NGC 5548 \rev  mass estimates by a factor of $\sim 3$, 
as they were significantly smaller then the other objects, and 
gave an excessive weight to these points.
 
\subsection{Weighting by Intrinsic Uncertainties}

We have used the standard procedure of weighting data with uncertainties 
(Bevington 1969, p. 118) which takes into account only uncertainties in one (the dependent)
variable. This weighting method may give too much weight to points with a very small
uncertainty in $M_{rev}$, since the uncertainty in $M_{ph}$ is always large,
as it has to account for the intrinsic uncertainties in the \ph method,
discussed above.  One could apply a 
``mixed weighting covariance''
method, where each data point is weighted considering its $x$ and $y$
uncertainties
(a detailed discussion of this subject can be found in Feigelson and Babu, 
1992).
However, also the \rev method has intrinsic uncertainties, as discussed in section 8 below. The errors quoted for $M_{rev}$ represent only the formal measurement errors, given by

\begin{equation}
\label{eq:dm}
\frac{\Delta M}{M} =
\left [ \left( \frac{\sigma_{\tau }}{\tau } \right)^2 +
        \left( \frac{2\sigma_{V}}{V} \right)^2 \right]^{1/2}
\end{equation}
Therefore also $M_{rev}$ has an additional intrinsic uncertainty (possibly larger than the measuring errors). Taking into account significant intrinsic 
uncertainties in both, $M_{rev}$ and $M_{ph}$ has the effect of giving all
points a similar weight, i.e. an unweighted fit.

\subsection{Interpreting the $M_{rev}-M_{ph}$ Relationship}

Our most important result is the very good correlation, consistent with a
linear relation, and the surprisingly low scatter.  We note that the outlying
objects in our sample are less than a factor of two away from the diagonal
($M_{ph}=M_{rev }$) line.

The slope ($b$) in the logarithmic fit deviating from unity may account for
intrinsic residual dependence of the \ph mass on on parameters such as
luminosity or line width, (for example, if one of the parameters in
eq.\ (\ref{eq:m7}), such as $U$ or $L_{ion}/L_v$ were dependent on $L$ or on
$M$). The fact that we find the slope to be consistent with unity implies
that there is no evidence for such an effect in our sample.
 
The constant ($a$) in the logarithmic fit gives an estimate of the calibration
coefficient, if $b$ is close to unity.  A more direct determination the
calibration coefficient is obtained by imposing $b=1$ and solving for the best
value of $a$, which we denote by $a_1$.  If $b$ is close to unity, this value
is given by 
\begin{equation}
\delta_M = < y > - < x >,
\end{equation}
where $< x>=\bar x$ and $<y>=\bar y$ are
the averages (weighted, Bevington p. 88, or unweighted, as discussed above).  For our sample we find 
$< {\lg (M_{rev}/\ms)}>=7.49$ and 7.64 while $<{\lg (M_{ph}/\ms)}>=7.29$ 
and $7.48$, for the
unweighted and weighted averages, respectively, so that $\delta_M= 0.20$ and 0.16,
respectively.

The difference $\delta_M$ is actually the empirical factor which best calibrates the
\ph method for determining the virial masses, so for the present sample and
our choice of parameters we find for the actual coefficient in eq. (\ref
{eq:m7}): $$f =10^ {\delta_M}\approx 1.6~ {\rm and} ~1.45,$$ respectively.

\section {The Radius Correlation}

The errors in the \ph virial mass determination 
(treated in more detail in the discussion section) consist of two main
ingredients: the uncertainties introduced by the \ph method of estimating the
BLR size and the uncertainty in matching the size derived to the velocity
dispersion implied by the line-width (and which 
measure of the line profile best describes the relevant velocity dispersion).
While these two problems are automatically solved in the \rev - rms method,
it important to understand their contribution to the errors in the \ph method.

By comparing the BLR size estimates of the \ph method with the reverberation
radii we can estimate which part of the scatter in the $M_{ph} - M_{rev}$
correlation (Fig.\ 2) 
is due to the uncertainties in the \ph model approximation
and the assumptions of average ionization parameter, density, and
$L_{ion}\approx L_{V}$.

The empirical result $r_{rev}\propto L^{1/2}$ (Kaspi \et 1997) would
predict a linear relation, $ \lg(r_{rev})=1.0 \lg (r_{ph})+C,$
as the \ph method in its simplest form implies $r_{ph}\propto L^{1/2}$.

The data and their interpretation are shown in figures 3a and 3b, respectively:
Figure 3a shows the lag vs.\ the monochromatic luminosity at 5100\AA.  
The logarithmic linear (unweighted) fit is 
${\lg \rm lag (days)} =(0.46\pm 0.1) \lg L_{44} +(1.50\pm 0.11)$
where $L_{44}=\lambda L_\lambda (5100\AA )/10^{44}\ers$.
Within the uncertainties this is consistent with the result of Kaspi \et (1997). Forcing an $R\propto L^{1/2}$ slope gives an almost identical result, 
$R_{H\beta}= 33 L_{44} ^{1/2}$light days,
or, expressed in terms of the luminosity integrated over the 0.1-1$\micron$
band rather than the monochromatic luminosity (assiming $L_\lambda\propto \lambda$ or $F_\nu\propto \nu^{-1}$)
$$R_{H\beta}= 15 L_{44,(0.1-1\micron )} ^{1/2}  {\rm light~ days} .$$

Figure 3b shows the lag vs.\ the BLR size estimated by the \ph method for
our sample.  The logarithmic linear fit is 
${\rm lag (days)} =0.92 r \mbox{\rm (lt-days)}+0.45$.
 The correlation ($r= 0.70$) 
is significantly less good than the correlation
of the masses.  This is caused by a few objects with large uncertainties being
well off the diagonal.  When the points are weighted by the lag
uncertainties, the slope becomes more deviant from unity --- $b=1.3\pm0.12$ ---
because several off-diagonal objects with small uncertainties dominate the
fit.  As discussed above, the unweighted fit may be more realistic as it
 approximately gives
the result of adding the intrinsic uncertainties. Efectively it softens the influence of the objects with very small lag error bars
in the weighting, lowering the slope back to be consistent with unity.

It is interesting to determine the calibration coefficient also for the
BLR size, derived from the \ph method. Considering figure 3b, 
the weighted averages are 
$<{\rm log(lag)}>= 1.22$ and 1.31
 and $<{\lg (r_{ph})}>= 0.81$ and $0.94$ for the unweighted and weighted 
methods, respectively.
The calibration factor $\delta_R=<{\rm log(lag)}>-<{\lg (r_{ph})}>$ is then 0.39
and 0.37, so that the radius calibration factor is 
\begin{equation}
f_R \approx 10^\delta_R = 2.4
\end{equation}
for both weighting methods. 
We see that the correlation between the BLR sizes given by the \rev and \ph 
methods is significantly weaker than the 
corresponding correlation between the masses and the scatter 
is larger.

\placefigure{fig3}

\section{Estimating $L_{ion}$ and $L/M$}
\subsection {The Ionizing Luminosity}
The  coefficient in eq. (\ref {eq:m7}) is 
$f= {3\over 4} f_L^{1/2}\bar {E_1}^{-1/2}$. 
We  may reverse this equation and use the value we have found for $f$  
in order to calculate the effective ratio $f_L$
between the visual luminosity 
$\lambda L_\lambda (5100\,{\rm \AA})$ and the ionizing 
luminosity:
$$f_L=\left({4\over 3}f\right)^2
\bar E_1\approx 1.8\times 10^{2\delta_M}\bar {E_1}\approx 4\bar {E_1},$$
However, the empirical
factor $f$ actually accounts for two components: the luminosity factor, $f_L$,
and the kinematic factor, $f_k$, which we have taken to be $3\over 4$ in both
methods, but it may contain also a possible factor contributed by the
mapping from the FWHM in the \ph method to the FWHM-RMS in the \rev method.

A more direct determination of $f_L$ can be obtained from the calibration 
factor of the BLR sizes:
since $R_{ph}\propto L_{obs}^{1/2}$ while the true \el~
 radius measured by the lag
is $R_{rev}\propto L_{ion}^{1/2}$, 
 we have $L_{ion}/L_{obs}\propto ( R_{rev}/R_{ph})^2$.
In our notation, eq.\ (\ref{eq:r}) gives
\begin{equation}
\label{equ:fl}
f_L =  Un_{10} \bar {E_1} \left ( {<R_{rev}>\over <R_{ph}>}\right ) ^2
\end{equation}

which for our choice, $Un_{10}=1$ becomes
$$f_L= 10^{2 \delta_R}\bar {E_1}= 6\bar {E_1}$$. 
For reasonable ionizing spectra
(Wandel 1997, Bechtold \et 1987)  $\bar E\sim 1.2-2$, so that 
that for our sample the ionizing continuum may be related to the monochromatic
luminosity at 5100\,\AA\ by
$$L_{ion}\approx 10 Un_{7-12}~ \lambda L_{\lambda}(5100{\rm \AA}).$$

For example, we can compare this empirical estimate of
the average photon energy to some published model estimates. 
Integrating the "Medium" AGN spectrum described in Bechtold et al. (1987)
gives a total bolometric luminosity of 1.4 times the value of
$\nu L_\nu$ at 1 Rydberg, which equals 7.7 times the value of
$\nu L_\nu$ at 5100 \AA .  The exact values are relatively insensitive
to the upper limit to the energy integral (the choice of hard-X-ray
cutoff).  About half of this total power is emitted
by the ``power-law" component longward of 1500 \AA .
For Bechtold et al.'s "Hard" AGN continuum energy distribution, a
somewhat smaller fraction of the total bolometric power emerges
at wavelengths longer than 1 Ryd, and the total bolometric power
is 13 times the value of $\nu L_\nu$ at 5100 \AA .
Within the uncertainties on average BLR parameters, either of
these models is consistent with the average AGN continuum spectrum
inferred from our comparison of photoionization and reverberation
estimates of BLR size.

\subsection {The Eddington Ratio}
We may use the high quality virial mass estimates obtained by the \rev -RMS
method for this sample, to estimate a mean $L/M$ relation. Combining this with
the above estimate of the ionizing luminosity (which presumably is comparable
to the total bolometric luminosity, as a large part of the AGN energy is
expected to be radiated in the UV) we may estimate the Eddington ratio.

Figure 4 shows the \rev - RMS virial mass vs.\ the 5100\,\AA\ 
luminosity for our sample.

\placefigure{fig4}

An unweighted linear fit gives $\lg (M/\ms ) = 0.54~ \lg (L_{44}) + 7.94$, with
a correlation coefficient $r = 0.68$. However, here the fit needs to be weighted only by the mass uncertainties, as
those of the luminosity are relatively small. A weighted fit yields 
\begin{equation}
\lg (M/\ms ) = (0.77\pm0.07) \lg (L_{44}) + (7.92\pm 0.04).
\end{equation}

We can calculate the \ed ratio from the difference of the weighted averages,
\begin{equation}
\delta_{ML}=<\lg (M/\ms )>-<\lg (L_{44})> = 8.0.
\end{equation}
Combining this with the definition of the Eddington luminosity  
$L_{Edd} = 1.3 \times 10^{38} ~(M/\ms )\,\ers $ we have for our sample
\begin{equation}
\lambda L_{\lambda}(5100\,{\rm \AA})\approx 0.01 L_{Edd}
\end{equation}
and 
\begin{equation}
L_{ion}/L_{Edd}\approx 0.1 Un_{10}.
\end{equation}

\subsection{Ionizing luminosities of individual objects}
We may apply eq. \ref{equ:fl}, (or inverse eq. (3)) to 
estimate the ionizing luminosity
of individual objects in our sample.
This gives
\begin{equation}
\label{equ:lion}
L_{ion}\approx \bar E_1 Un_{10}\lambda L_{\lambda}(5100\,{\rm \AA})
\left ( {R_{rev}\over R_{ph}}\right ) ^2 = 6\times 10^{43} E_1 Un_{10}
\left ( {{\tau}\over 10{\rm days}}\right ) ^2 \ers ,
\end{equation}
where $\tau$ is the lag (the centroid of the cross-correlation function). 
Since the ionizing luminosity is likely to be a large fraction of the bolometric luminosity, and the virial \rev -rms mass is our most reliable estimate of the
 mass of the central \bh , these values give a direct measure of the true 
Eddington ratio
$$L/L_{Edd}\approx L_{ion}/M_{rev} .$$ 
The last two columns in table 2 show the ionizing luminosity and total 
\ed ratio of the objects in our sample.

\placefigure {fig5}

Fig. 5 shows the virial (reverberation) mass vs. the ionizing luminosity
for the objects in our sample. We note that most objects in our sample have 
$L/L{Edd}\sim 0.01-0.3$.
We also note that there is a strong trend of the Eddington
ratio to increase with luminosity:  
an unweighted fit to the Eddington ratio vs. luminosity gives 
$$L/L_{Edd}= 0.08 L_{44}^{0.7},$$
where $L_{44}$ is the luminosity at 5100\AA .
However, since $L_{ion}\propto \tau^2$ (eq. \ref{equ:lion})
and $M\propto \tau \times (FWHM)^2$, this correlation may be a mere reflection
of the correlation of the lag with luminosity.

\section {Discussion}

\subsection{Results}
From the present sample we find that the virial masses of the 19 AGN in our
sample estimated by the \ph method are very well correlated with the more
direct estimates from the \rev - RMS method.  We find the relation is
consistent with a linear one, that is, $M_{rev}\propto M_{ph}$.  For an
average value of $Un_e=10^{10}$ and assuming $L_{ion}=f_L\lambda F_\lambda 
(5100\,{\rm \AA})$
we find a calibration factor of $f\approx 1.5$ 
where $M_{rev}=f M_{ph}$, which corresponds to $f_L\approx 10$.

The fact that the \ph masses are better correlated with the \rev - RMS
masses than the corresponding BLR size with the \rev distances
 contradicts the intuitive  prediction that the
scatter in the masses should be larger, due to the additional uncertainty in
translating line profiles to the appropriate velocity dispersion.
This may indicate that there is an internal compensation mechanism, which 
reduces scatter in the \ph mass determination.
This may be related to the difference between the \rev and the \ph methods:
the former measures the size of the variable component of the BLR,
while the latter measures the emissivity-weighted size, and the two may differ.
However, when applied to the virial mass, both methods measure the same mass.

This conclusion is strongly supported by the variability of NGC 5548 (Fig.\
1), where the variations in the lag of the line width are significantly larger
than the variations in the combination lag$\times \vFWHM^2$. (The latter is
actually consistent with remaining constant, within the uncertainties).

Also, examining the deviations of individual objects from the linear relation
in the mass and size plots, we note that several objects which deviate
significantly in Fig.\ 4 (the BLR size correlation) have smaller deviations in
the mass correlation (Fig.\ 3).

We have also estimated the ionizing luminosity, the mass-luminosity
relation and the Eddington ratio for our sample.
Applying the calibrated \ph method to many AGN 
may reveal a trend for these important parameters in various groups of AGN.

Obtaining reliable \bh~ masses for many AGN may also provide important
insight for the relation between AGNs and their host galaxies, as has already
been demonstrated for the present sample (Wandel 1999).
\subsection { Comparison with Accretion-Disk Models}

AGN masses may be estimated also by fitting the continuum optical-
UV spectrum by a thin \ad model spectrum (e.g. Malkan 1990; 
Wandel and Petrosian 1988; Laor 1990).
We have compared the reverberation and photoionization mass
estimates with those obtained by standard thin-accretion disk
fitting of the continuum energy distribution.  Specifically,
more than half of the calibrating Seyfert 1's have been 
fitted in Sun and Malkan (1989), Malkan (1990), and 
Alloin et al. (1995).
The black hole masses estimated from continuum fitting
are generally correlated with those derived from
the emission lines.  
We find that in general the \bh~ mass obtained from the \ad method tends to be
larger than the mass found from the \rev data.
In order for the normalizations to be consistent with the \rev masses the 
parameters of the disk models have to be 
chosen to give the smallest possible black hole masses.
Specifically, the disk fits must all assume non-rotating
(Schwarzschild) black holes, and zero inclination 
(face on) to give a normalization consistent with the emission line
estimates.  
For most of the objects in our sample, inclined disks, or rotating 
(Kerr) black holes
would yield larger continuum-fitting black hole masses,
inconsistent with the emission line fits. 
While for Kerr \bh \ad models the \bh~mass required to fit a given 
continuum depends sensitively on the inclination, 
the dependence in the \sw case is much less steep. 
For a \sw \bh~ one may therefore soften the constraint of small
inclination, taking into account the intrinsic uncertainties in the
mass determination and in the \ad-fitting method.
This argument and the constraint of low \bh~ masses seem to infer that 
\bhs~ in AGN (at least in the Seyfert 1 galaxies in our sample) tend
to have little or no rotation, being near-\sw .
This subject will be elaborated in a separate work.

\subsection {Selection effects}

A major question that comes up when trying to apply the calibration result 
obtained from this sample to other AGNs without is whether the sample is
representative. The objects in our sample - AGNs with adequate
reverberation data - were selected by the criteria that make them "promising"
for \rev mapping: evidence or good chances to observe continuum and \el
variability. In most cases this meant some previous indication that the 
\el s varied.
The evident selection criterion is therefore variability, in particular
\el~ variability. However, variability is a feature common to many if not 
most AGNs, and it is difficult to identify an obvious way how such
 selection criteria can bias our results. If, for example, 
the more variable sources had
systematically broader mean H$_\beta$ line profiles, then the \bh~ masses
derived using the coefficient from our sample would be underestimated.

However, the line width enters in the same way in both mass estimation methods,
only it is the mean profile in the photoionization method and the rms profile
in the reverberation one, so it is the difference between the mean and rms
profile width which should be considered, and that one shows no trend with
luminosity (Fig. 6). We also note that the FWHM difference shows no preferred 
direction: it is positive for some objects and negative for others.

\placefigure{fig6}

Another question that may be posed is wheather the selection on objects on the
basis of \el -variability introduces a bias at the parameter-space coverage:
wheather our sample is representative of the AGN population. We have plotted
our sample together with  large samples of PG quasars 
and Seyfert 1 galaxies (both, narrow and broad line Seyferts) in the luminosity
vs. H$\beta$ line-width
plane (Fig. 7). We can see that our sample provides a fair representation of the
local AGN population, except of the high luminosity end. This limitation 
can be expected, as the BLR size is correlated with luminosity, so very luminous
objects are likely to have larger BLRs, and hence longer response times, hence
they would be more
difficult for reverberation measurement. However, as Fig. 7 reveals no 
obvious correlation between luminosity and line width, this restiction is not
likely to bias our results.

\placefigure{fig7}
\subsection {Uncertainties of the \rev method}

Although the BLR sizes determined from the lag are much more reliable  than those given by the \ph method, which involves much larger intrinsic uncertainties (see below), it has uncertain factors depending on the geometry
of the emitting gas:

\begin {itemize}
\item
What is the error introduced by using moments of the transfer function
rather than the full function?
The shape of
the cross-correlation function depends not only on the transfer function,
but also on the form of the continuum variations, so one can't read
much from it. However, its first moment (the centroid) is rather less
dependent on the form of the continuum variations. To REALLY address the
problem correctly, one needs the transfer functions, but accurate transfer
recovery is somewhat beyond the scope of what can be done with the data
we have at present.

\item
How is the lag time related to the geometrical distance of the line emitting
gas from the illuminating continuum, and how does it depend on the geometry,
in particular of a radially extended distribution?
Also here there is a dependence on the continuum variations.
For example, if the emitting region has an extended range of radii, 
$R_{in}--R_{out}$ the outer parts will not 
be able to respond to continuum changes with a timescale sorter than
$\sim R_{out}/c$, while the inner parts will.
In that case a more adequate measure of the size may be obtained using
the peak rater than the centroid of the cross-correlation function.
\end{itemize}

 More uncertainties are introduced by combining the
lag with the line profile to give a mass estimate:

\begin {itemize}
\item
What measure of the line profile gives the appropriate velocity dispersion
to combine with the lag? This becomes more complicated when the line emitting
gas has a radially extended geometry, as the centroid lag, which is the 
responsivity weighted delay, may be differently weighted than the FWHM,
which is the emissivity weighted line-of-sight velocity dispersion.

\item
Does the width of line profile measure the velocity dispersion in the same
gas the distance of which is measured by the lag? Using the rms rather than
the mean profile reduces, but does not eliminate this uncertainty. 
\item
What is the relationship between the emissivity-weighted line-of-sight velocity
measured by the line width and the central mass? This would depend on the shape
of the orbits, which may introduce a systematic variation of a factor of a few.
\item
What happens if the illuminating radiation does not originate directly from
the central mass? In that case the lag and the line width would refer to 
different parts, the combination of which to give a virial mass estimate would
 depend on the specific geometry.
\end {itemize}
All the above uncertainties are systematic, and introduce an error factor of
 a few, significantly larger than the random errors quoted in table 1.
However, being systematic uncertainties, they will apply to all objects alike,
altering the calibration coefficient but not effecting
the linear correlation between the masses derived by the two methods. 
However, if there were luminosity-dependent effects, for example if the
BLR geometry would change with $L$, or if the relation between the rms and
the mean line width depended on $L$ - of on $M$ - this would change the slope
of the correlation.
As seen from Fig 6, such a systematic difference between the rms and mean
FWHM in our sample is not present.
\subsection {Uncertainties of the \ph method}
The basic \ph method applied above to check the feasibility of the
calibration makes several simplifying assumptions, which need to be
examined in more detail or replaced by a more elaborate treatment.
\begin{itemize}
\item
Potential ambiguities arise as a result of the radial extent and
ionization stratification of the BLR.  The effect of these complications may
be studied within the framework of the locally maximally emitting cloud model
(Baldwin \et 1995; Korista \et 1995). 
\item
The physical parameters used to calculate the distance (density and 
ionization parameter )
may vary from object to object and differ also for different lines in the same
object. The adjustable factor in the basic \ph method  
depends on the values used
for the ``average'' ionization parameter, density, and $L/L_{ion}$. 
It is possible to use spectral data on different emission lines, 
to estimate the values of the ionization parameter and density for 
individual objects.
\item
The mapping from the line width to the actual virial velocity depends on the distribution of the line emission in physical and velocity space, and
how this is related to the line-width measure (e.g., FWHM, FWZI, intermediate
width, skewness). The convolution of extended-emission
BLR models with various possible 
dynamics of the BLR gas may be used to find an improved
relation between the line profile and the effective velocity dispersion in the
line emitting gas, induced by the central mass.
\item
In extended BLR geometries, also 
 the physical characteristics (such as density, differential covering
area, emissivity) are likely to vary with radius, as indicated e.g. by two-zone photoionization modeling of several
line ratios, which suggests a stratified BLR possibly extending into an
"intermediate line region" (Brotherton \et 1994).
\end{itemize}
Compensating for these potential ambiguities in the photoionization method
may give an improved \ph method of virial mass determination,
and may also reduce the systematic error sources in the \ph  mass
estimates and the scatter in the $M_{ph}$vs. $M_{rev}$ diagram.

Comparing masses and BLR sizes obtained by improved \ph models to the 
\rev - RMS values may in turn be used to test the \ph BLR models.

\section {Summary}
We have estimated the virial mass of 19 AGN from the BLR size and emission-line
width, using two different methods: the accurate \rev mapping combined with the
variable part of the emission-line, and the \ph method, combined with the plain
line width. We have shown that the two mass estimates for all objects in our
sample are highly correlated, and consistent with a linear relation, hence
we were able to calibrate the \ph method of mass and BLR size estimate.
This should make possible mass estimates for virtually all AGN with H$\beta$
line width measurements, which will be presented in a subsequent work.
Our sample significantly enhances the available AGN central mass estimates: 
it has twice the number of objects previously available, it has better
mass values, using the rms rather than the mean line profile, and all objects
 are uniformly treated.
The calibration also enables us to estimate the ionizing luminosity, which
cannot be directly observed. For our sample we find it to be on average 
10 times the
 monochromatic luminosity at 5100\AA . Combining the ionizing luminosities
with the \rev mass estimates our sample gives a
mass-luminosity relation of $M\propto L^{0.8}$, with an  
Eddington ratio distributed in the range 0.01--0.3.

\acknowledgments

We are grateful to Shai Kaspi and Hagai Netzer, for permission to use the Wise
Observatory data prior to publication, and to the anonymous referee for 
valuable suggestions.
AW acknowledges the hospitality of the Astronomy Department at UCLA.
We are grateful for support of this work by the National Science Foundation
through grant AST--9420080 to The Ohio State University.


\clearpage

\figcaption[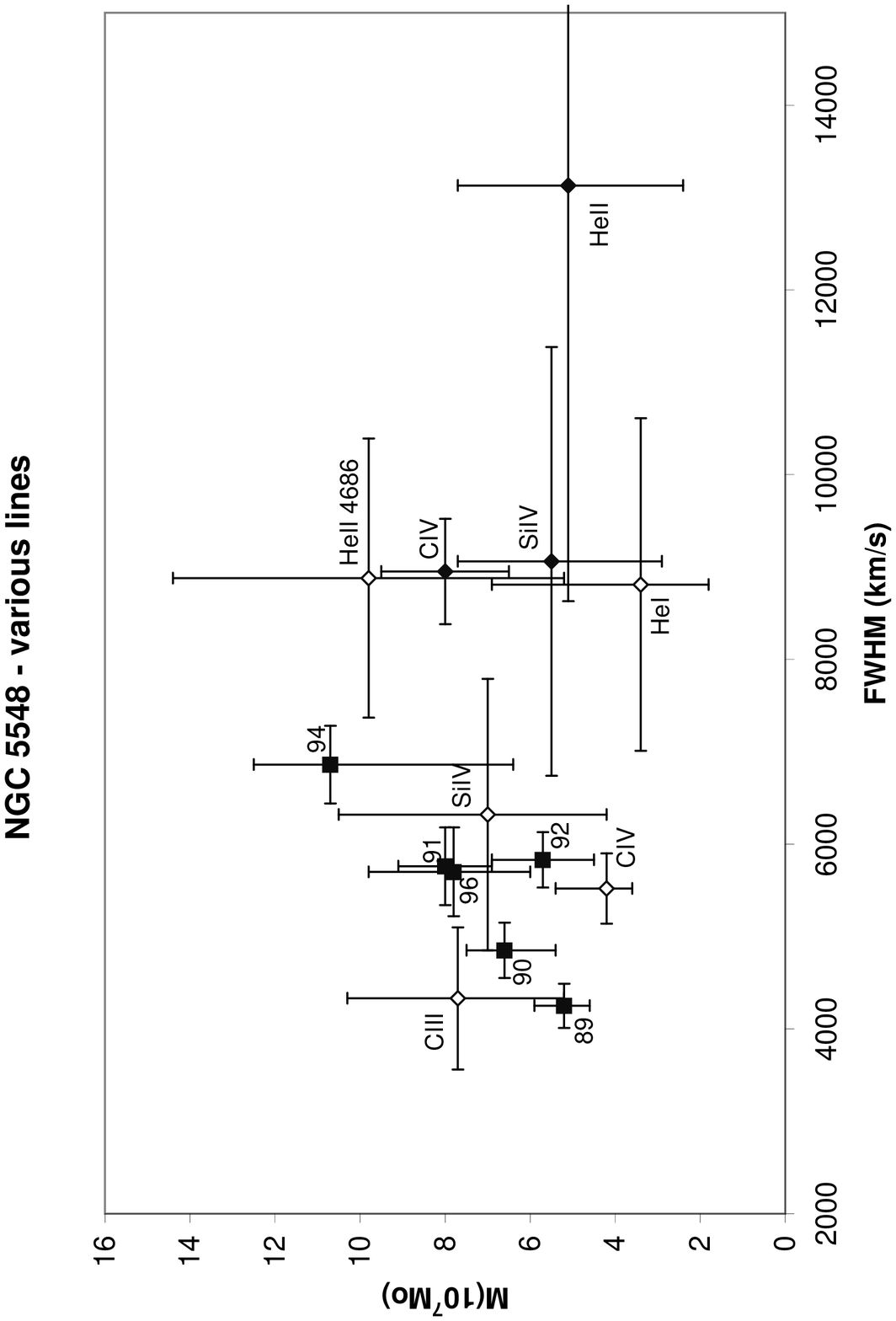]{The virial mass of NGC 5548 vs.\ FWHM-rms for H$\beta$ (6 observing seasons
in the years 1989-1996) and five other lines in the years 1989 and 1993 
 (see text). Data taken from table 1 of Peterson and Wandel (1999).
\label{fig1}}

\figcaption[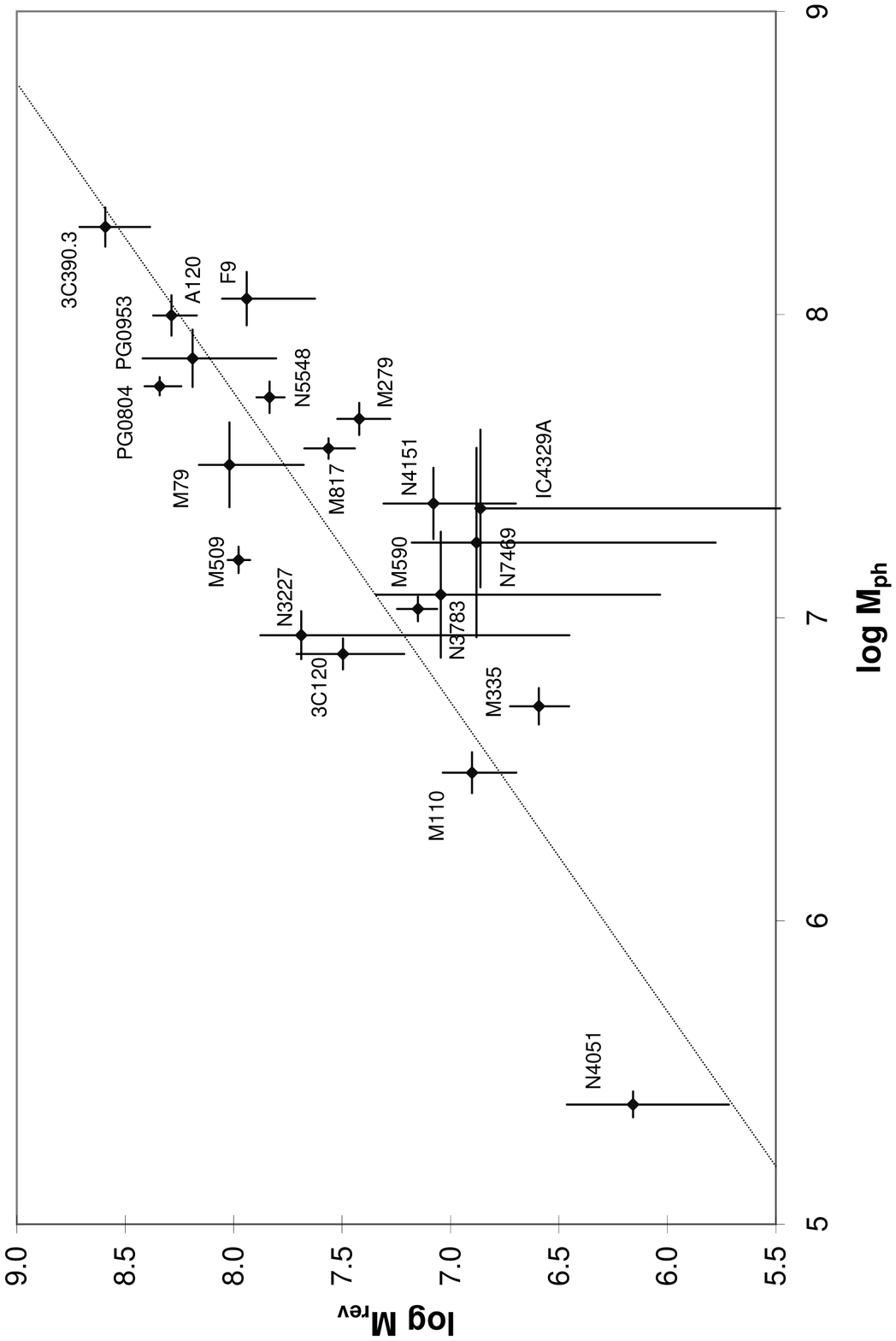]{ Virial mass  obtained
from the reverberation-rms data vs. the virial mass estimated by the \ph method.
The dotted line shows the best linear ($M_{rev}\propto M_{ph}$) fit  weighted by the uncertainty in $M_{rev}$ for each object. \label{fig2}}

\figcaption[fig3.ps]
{a. Reverberation centroid lags vs monochromatic luminosity at 5100\AA.
b.\ Reverberation radius vs. BLR size estimated with the \ph method. 
The dotted line shows the best linear ($\tau \propto L^{1/2}$ or
$R_{rev}\propto R_{ph}$) fit  
weighted by the uncertainties in the lag. Formally, the uncertainties in the
\ph estimates are the uncertainties in $L(5100\AA )$ which are very small.
The intrinsic uncertainties in the \ph estimates are not shown.\label{fig3}}

\figcaption[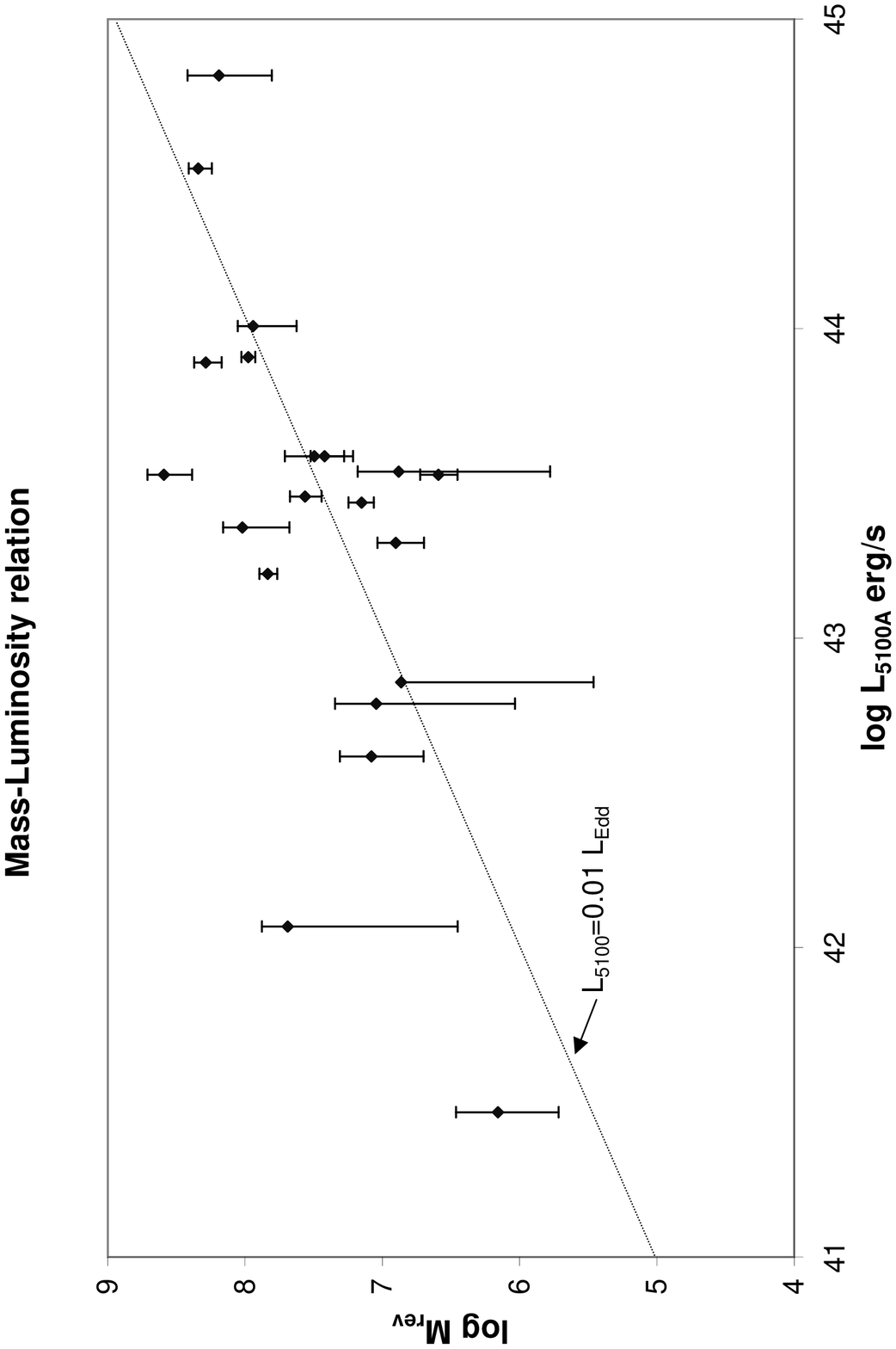]
{Reverberation-RMS mass vs. monochromatic luminosity. The dotted line shows the best linear ($M_{rev}\propto L$) fit  weighted by the 
uncertainties in $M_{rev}$.\label{fig4}}

\figcaption[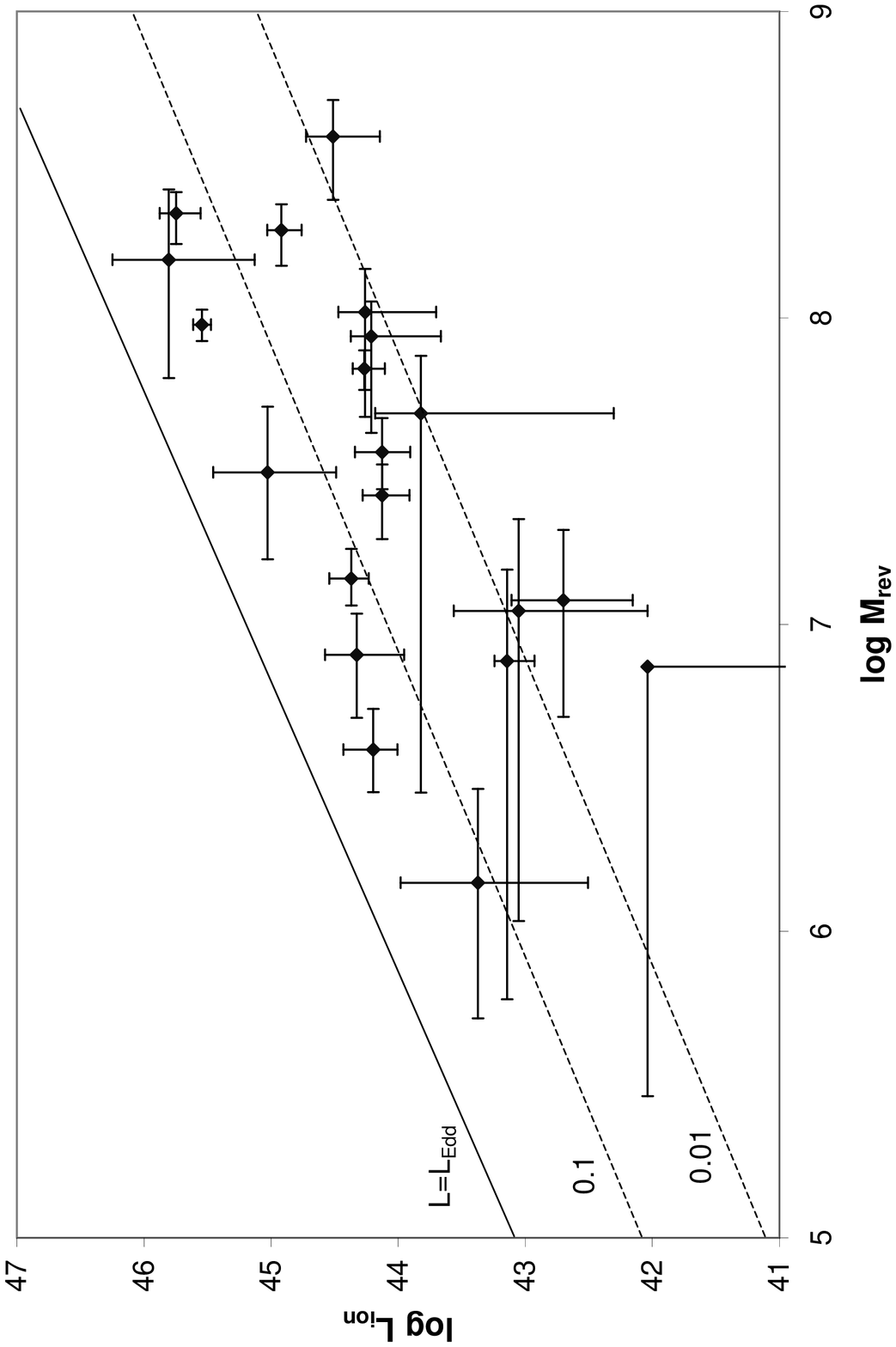]
{Ionizing luminosity (derived from the lag, see text) vs. reverberation-RMS mass. The diagonal lines correspond to an Eddington ratio of 1 and 0.01.
\label{fig5}}

\figcaption[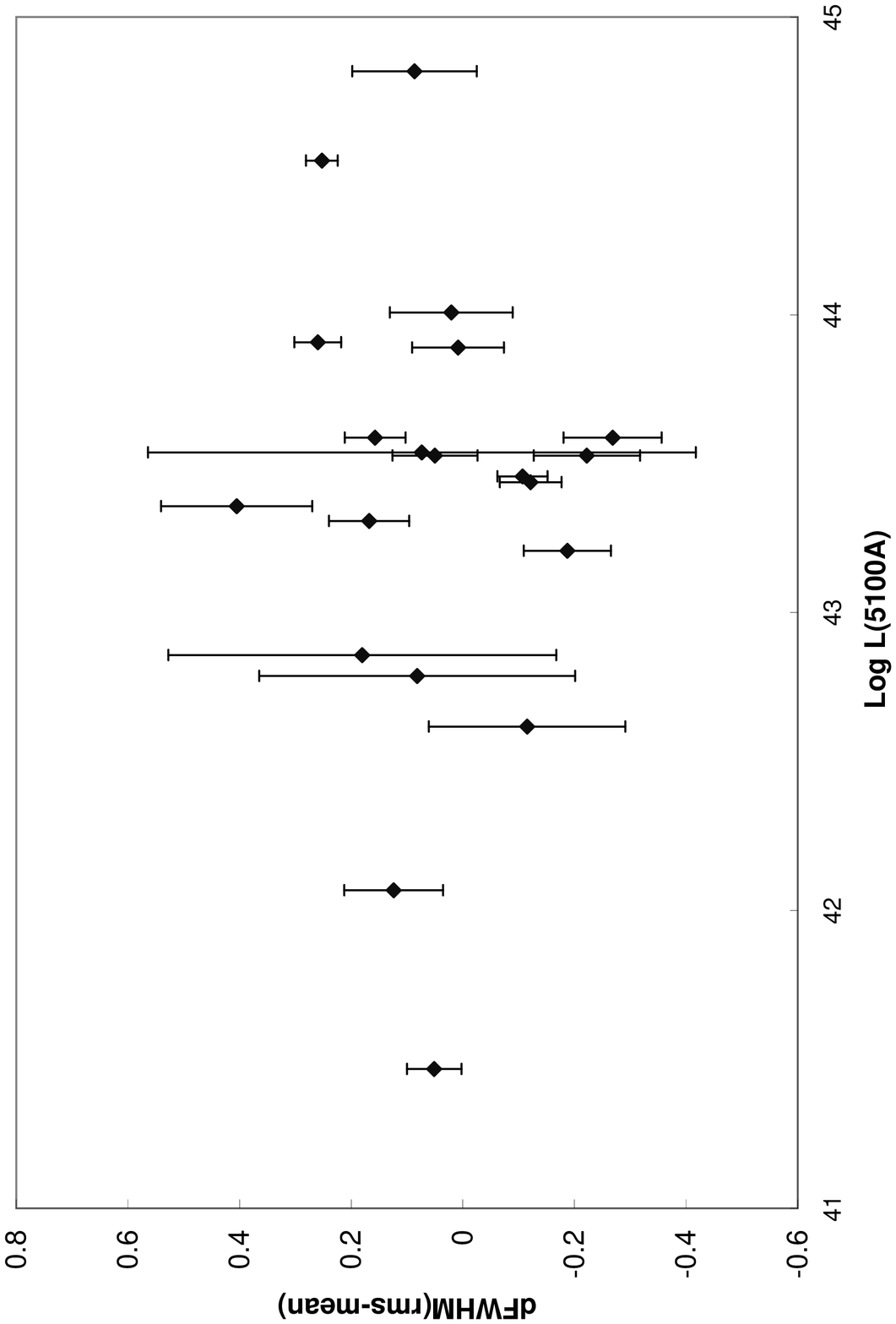]
{The fractional difference between the rms and mean FWHM(H$\beta$) line
profiles (FWHM difference divided by FWHM(rms) vs. luminosity. \label{fig6}}

\figcaption[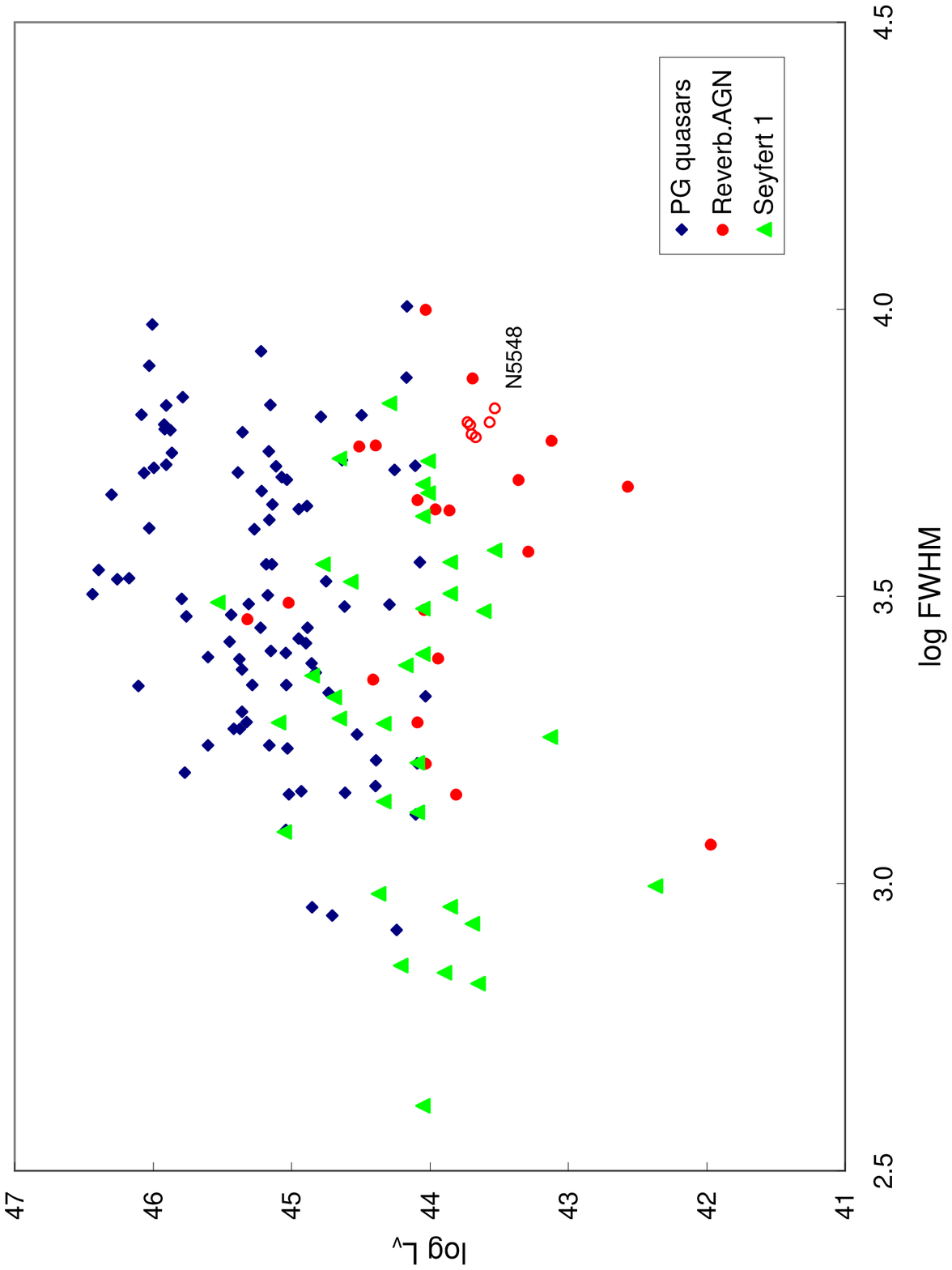]
{The reverberation AGN sample (circles) in the visual luminosity vs. FWHM(H$\beta$) plane  
compared to other AGN: PG quasars (diamonds; from Boroson and Green, 1994) and Seyfert 1
galaxies (triangles; from Boller, Brandt and Fink, 1996). Open circles indicate the six
years of data for NGC 5548. \label{fig7}}

 \begin{table}
\caption{Continuum Luminosity, FWHM H$\beta$ (mean and rms), and 
Continuum-Line Lags \label{tbl-1}}
\begin{center}
\begin{tabular}{lllll}
\tableline
{}&{}&{}&{}&{}\\
Name & $\log \lambda L_\lambda $(5100\AA )& 
	\vFWHM (mean)&\vFWHM (rms) &  lag $\tau_{cent}$ \\ 
{}&{(\ers)}&{(\kms)}&{(\kms)}&{(days)}\\
{}&{}&{}&{}&{}\\
\tableline
{}&{}&{}&{}&{}\\
3C\,120     &     43.59&    1910 &$ 2210 \pm 120$&  $ 43.8^{+27.7}_{-20.3} $\\
3C\,390.3   &     43.53&   10000 &$10500 \pm 800$&  $ 24.2^{+ 6.7}_{- 8.4} $\\
Akn\,120    &     43.89&    5800 &$ 5850 \pm 480$&  $ 38.6^{+ 5.3}_{- 6.5} $\\
F\,9        &     44.01&    5780 &$ 5900 \pm 650$&  $ 17.1^{+ 3.5}_{- 8.0} $\\
IC\,4329A   &     42.86&    5050 &$ 5960 \pm2070$&  $  1.4^{+ 3.4}_{- 2.9} $\\
Mrk\,79     &     43.36&    4470 &$ 6280 \pm850$&  $ 18.1^{+ 4.9}_{- 8.6} $\\
Mrk\,110    &     43.31&    1430 &$ 1670 \pm 120$&  $ 19.5^{+ 6.5}_{- 6.8} $\\
Mrk\,335    &     43.53&    1620 &$ 1260 \pm 120$&  $ 16.8^{+ 5.2}_{- 3.3} $\\
Mrk\,509    &     43.91&    2270 &$ 2860 \pm 120$&  $ 79.3^{+ 6.5}_{- 6.2} $\\
Mrk\,590    &     43.44&    2470 &$ 2170 \pm 120$&  $ 20.5^{+ 4.5}_{- 3.0} $\\
Mrk\,817    &     43.46&    4490 &$ 4010 \pm 180$&  $ 15.5^{+ 4.3}_{- 3.5} $\\
NGC\,3227   &     42.07&    4920 &$ 5530 \pm 490$&  $ 10.9^{+ 5.6}_{-10.9} $\\
NGC\,3783   &     42.79&    3790 &$ 4100 \pm1160$&  $  4.5^{+ 3.6}_{- 3.1} $\\
NGC\,4051   &     41.47&    1170 &$ 1230 \pm  60$&  $  6.5^{+ 6.6}_{- 4.1} $\\
NGC\,4151   &     42.62&    5910 &$ 5230 \pm 920$&  $  3.0^{+ 1.8}_{- 1.4} $\\
NGC\,5548\tablenotemark{a}
            &     43.21&  6300 &$ 5500 \pm 400$&  $ 21.6^{+ 2.4}_{- 0.7} $ \\
NGC\,7469\tablenotemark{b}   
  	  &     43.54&    3000 &$ 3220 \pm1580$&  $  5.0^{+ 0.6}_{- 1.1} $\\
PG\,0804+762&     44.52&    3090 &$ 3870 \pm 110$&  $100.0^{+16.3}_{- 19.8} $\\
PG\,0953+414&     44.82&    2890 &$ 3140 \pm 350$&  $107.1^{+71.2}_{- 58.0} $\\
{}&{}&{}&{}&{}\\
\tableline
\end{tabular}
\tablenotetext{a}{For NGC\,5548, the lag and the FWHM(rms) are derived from
the whole data set of the years 1989--1996.}

\tablenotetext{b}{For NGC\,7469, the cross-correlation lags are
relative to the ultraviolet continnuum variations at 1315\,\AA\
on account of the wavelength-dependent continuum time delays
(see Wanders et al.\ 1997; Collier et al.\ 1998; 
Peterson et al.\ 1998b).}
\end{center}
\end{table}
\newpage
\begin{table}
\caption{Reverberation BLR Sizes and Central Masses, Compared with 
Photoionization  Sizes and Masses.
The last two columns give the ionizing luminosity 
(derived from the lag, see text) and the corresponding Eddington ratio}
 \label{tbl-2}
\begin{center}
\begin{tabular}{llllllll}
\tableline
{}&{}&{}&{}&{}&{}&{}&{}\\
Name   &  log $R_{ph}$&  log lag& log $M_{ph}$&  log $M_{rev}$ & $M_{rev} (10^7 \ms)$&log $L_{ion}$&log$(L_{ion}/L_{Edd})$ \\
{}&{}&{}&{}&{}&{}&{}&{}\\
\tableline
{}&{}&{}&{}&{}&{}&{}&{}\\

3C\,120     &$  0.92$& 1.64& 6.86& 7.49  &  $ 3.1^{+2.0}_{-1.5}$&45.03&-0.57\\
3C\,390.3   &$  0.89$& 1.38& 8.26& 8.59  &  $39.1^{+12}_{-15}$&44.51&-2.19 \\
Akn\,120    &$  1.07$& 1.59& 7.97& 8.29  &  $19.3^{+4.1}_{-4.6}$&44.92&-1.48 \\
F\,9        &$  1.13$& 1.23& 8.03& 7.94  &  $ 8.7^{+2.6}_{-4.5}$&44.21&-1.84 \\
IC\,4329A   &$  <0.56$& 0.15& 7.34& $<6.86 $ &  $<0.73$   &$<42.04$& $<-2.93$\\
Mrk\,79     &$  0.81$& 1.26& 7.48& 8.02  &  $10.5^{+4.0}_{-5.7}$ &44.26&-1.87\\
Mrk\,110    &$  0.78$& 1.29& 6.46& 6.91  &  $0.80^{+0.29}_{-0.30}$ &44.33&-0.69\\
Mrk\,335    &$  0.89$& 1.23& 6.68& 6.58  &  $0.39^{+0.14}_{-0.11}$&44.20&-0.49 \\
Mrk\,509    &$  1.08$& 1.90& 7.17& 7.98  &  $9.5^{+1.1}_{-1.1}$&45.54&-0.54 \\
Mrk\,590    &$  0.85$& 1.31& 7.00& 7.15  &  $1.4^{+0.3}_{-0.3}$&44.37&-0.89 \\
Mrk\,817    &$  0.86$& 1.19& 7.53& 7.56  &  $3.7^{+1.1}_{-0.9}$&44.13&-1.54 \\
NGC\,3227   &$  0.16$& 1.04& 6.92& 7.69  &  $4.9^{+2.7}_{-5.0}$&43.82&-1.98 \\
NGC\,3783   &$  0.52$& 0.65& 7.05& 7.04  &  $1.1^{+1.1}_{-1.0}$&43.05&-2.10 \\
NGC\,4051   &$ -0.14$& 0.81& 5.37& 6.15  &  $0.14^{+ 0.15}_{-0.09}$& 41.57&-0.84\\
NGC\,4151   &$  0.44$& 0.48& 7.35& 7.08  &  $1.2^{+ 0.8}_{-0.7}$&42.70&-2.49 \\
NGC\,5548 &$  0.73$& 1.26& 7.70& 7.83  &  $6.8^{+ 1.5}_{-1.0}$&44.27&-1.83 \\
NGC\,7469   &$  0.90$& 0.70& 6.87& 6.88  &  $0.76^{+ 0.75}_{-0.76}$&43.14&-1.86 \\
PG\,0804+762&$  1.39$& 2.00& 7.74& 8.34  &  $21.9^{+ 3.8}_{-4.5}$&45.75&-0.70 \\
PG\,0953+414&$  1.54$& 2.03& 7.83& 8.19  &  $15.5^{+ 10.8}_{-9.1}$&45.81&-0.49 \\
{}&{}&{}&{}&{}&{}&{}&{}\\
\tableline
\end{tabular}
\end{center}
\end{table}

\begin{table}
\caption{Reduced $\chi^2$ and probabilities (in parenthesis) for the three fitting schemes, constant 
mass (columns 3-4), lag$\propto (FWHM)^2$ (y=-2x+a, columns 5-6), and the best fit (y=-bx+a,
columns 7-9). The first column indicates the data set, the second one (n) - the number of points.
}
 \label{tbl-3}
\begin{center}
\begin{tabular}{lcccccccc}
\tableline
{}&{}&{}&{}&{}&{}&{}&{}&{}\\
Data set   &  $n$ & $\bar M_7$ & $\chi^2(\bar M)$ & a & $\chi^2(-2x+a)$ & $b$ & $a$ & $\chi^2(-bx+a)$\\
&(2)&(3)&(4)&(5)&(6)&(7)&(8)&(9)\\
{}&{}&{}&{}&{}&{}&{}&{}&{}\\
\tableline
{}&{}&{}&{}&{}&{}&{}&{}&{}\\
All lines, H$\beta~$89-96&14& 6.1&1.09(0.35)& 8.85 & 2.10(0.04) & -2.05$\pm$0.13 & 8.85$\pm$ 0.49 & 1.78(0.05)\\
All lines, H$\beta~$89   &9 & 5.5&0.92(0.49)& 8.58 & 1.57(0.15) & -1.98$\pm$0.13 & 8.50$\pm$ 0.51 & 1.22(0.29)\\
All lines, H$\beta~$ average&9&6.0&0.87(0.51)& 8.62 & 1.44(0.19) & -2.22$\pm$0.18 & 9.51$\pm$ 0.69 & 0.96(0.45)\\
H$\beta$ 89-96, centroid &6 & 6.6&1.72(0.15)& 8.66 & 3.51(0.01) & -0.76$\pm$0.39 & 4.06$\pm$ 1.43 & 0.93(0.45)\\
H$\beta$ 89-96, peak	 &6 & 6.6&0.58(0.68)& 8.66 & 1.67(0.15) & -1.47$\pm$0.27 & 6.67$\pm$ 1.00 & 0.66(0.62)\\
{}&{}&{}&{}&{}&{}&{}&{}&{}\\
\tableline
\end{tabular}

\end{center}
\end{table}

\end {document}